\documentclass[aps, prd, amsmath, floats, floatfix, twocolumn, superscriptaddress, nofootinbib, showpacs,]{revtex4}
\usepackage[switch]{lineno}

\usepackage{hyperref}
\usepackage{amssymb}
\usepackage{amsmath}
\usepackage{verbatim}
\usepackage{mathrsfs}
\usepackage{amsfonts}
\usepackage{latexsym}
\usepackage{epsfig}
\usepackage{epstopdf}
\usepackage[usenames, dvipsnames]{color}
\usepackage{hyperref}
\hypersetup{
    colorlinks=true,
    linkcolor=blue,
    citecolor=blue,
    urlcolor=red,
    pdftitle={Fermi bubbles in SFDM},
    }

\definecolor{pink}{rgb}{0.9,0,0.6} 

\newcommand{\beq}{\begin{equation}}
\newcommand{\eeq}{\end{equation}}
\newcommand{\bea}{\begin{eqnarray}}
\newcommand{\eea}{\end{eqnarray}}
\newcommand{\beqn}{\begin{equation*}}
\newcommand{\eeqn}{\end{equation*}}
\newcommand{\bean}{\begin{eqnarray*}}
\newcommand{\eean}{\end{eqnarray*}}

\long\def\symbolfootnote[#1]#2{\begingroup%
\def\thefootnote{\fnsymbol{footnote}}\footnote[#1]{#2}\endgroup}

\begin{document}


\title{Alternative explanation of Fermi Bubbles using Scalar Field Dark Matter} 

\author{Tonatiuh Matos}\email[0000-0002-0570-7246 ]{tonatiuh.matos@cinvestav.mx}
\affiliation{Departamento de F\'isica, Centro de Investigaci\'on y de Estudios Avanzados del IPN, A.P. 14-740, 07000 CDMX.,  M\'exico.}

\author{Bryan Mendoza-Meza}\email{luis.mendoza@cinvestav.mx}
\affiliation{Departamento de F\'isica, Centro de Investigaci\'on y de Estudios Avanzados del IPN, A.P. 14-740, 07000 CDMX.,  M\'exico.}

\author{Leonardo San.-Hernandez}\email[]{leonardo.sanchez@cinvestav.mx}
    \affiliation{Departamento de F\'isica, Universidad Aut\'onoma Metropolitana Iztapalapa, San Rafael Atlixco 186, CP 09340, Ciudad de M\'exico,
  M\'exico.}

  \author{Abdel Perez-Lorenzana}
\email[0000-0001-9442-3538 ]{aplorenz@fis.cinvestav.mx}
\affiliation{Departamento de F\'isica, Centro de Investigaci\'on y de Estudios Avanzados del IPN, A.P. 14-740, 07000 CDMX.,  M\'exico.}

\author{Maribel Hern\'andez-M\'arquez}\email[]{maribel.hernandez@nucleares.unam.mx}
\affiliation{Instituto de Ciencias Nucleares, Universidad Nacional Aut\'onoma de M\'exico, Circuito Exterior C.U., A.P. 70-543, Ciudad de M\'exico 04510, M\'exico}


\date{\today}


\begin{abstract}
In recent times, the Scalar Field Dark Matter (SFDM) model (also called Fuzzy, Wave, Ultralight dark matter model) has received much attention due to its success in describing dark matter on both cosmological and galactic scales. Several challenges of the Cold Dark Matter (CDM) model can be explained very easily and naturally by the SFDM model. Two of these challenges are to describe the anomalous trajectories of satellite galaxies called the Vast Polar Structure (VPOS) and to explain the magnetic fields observed in our galaxy. In previous works an alternative explanation for VPOS and the magnetic fields of our galaxy was proposed using the SFDM excited states, explaining the anomalous trajectories in a natural and simple way. In this work we use the same dark matter endowed with a extremely small charge that explains the magnetic fields of our galaxy to show that these excited states of SFDM can provide a very simple and natural explanation for Fermi Bubbles (FBs). If this assumption is correct, we should see FBs in several more galaxies, these observations would take place in the near future and could be crucial to the ultimate answer to the nature of dark matter.
\end{abstract}




\maketitle


\label{sec:introduction}
\textbf{Introduction:} The nature of Dark Matter (DM) remains one of the most challenging open questions in modern cosmology \cite{Arbey:2021gdg}. The $\Lambda$ Cold Dark Matter ($\Lambda$CDM) model has been able to explain all the cosmological observations we have so far, maybe with the exception of the so-called $H_0$ tension \cite{delaMacorra:2021hoh}. Nevertheless, at galactic scales, this model has several strong challenges to explain, as some of the observed features of the galaxies are not sufficiently well explained by the model or the needed physics complicates too much the model \citep{cdm_challenges, cdm_challenges2}. This is why people look for alternatives that can solve these problems in a more convincing way. One of the alternatives that do so is the Scalar Field Dark Matter (SFDM) model, which explains the universe at cosmological scales with the same accuracy as the $\Lambda$CDM model \citep{Matos:1998vk,sf40,sf3}, {but is able to explain small-scale structure with much simpler physics in a natural way \cite{Matos:2023usa}.}
 The model has been rediscovered several times as Fuzzy \citep{sf2}, Wave \citep{Bray:2010fc,sc4}, BEC DM \citep{sf10}, and more recently as Ultra-light DM  \citep{Hui:2016ltb}.
The SFDM model has still a long history, there are some isolated works in the 90's of the past century \cite{ sf80,sf8}, but the systematic study of this model began at the end of the 90's with the study of galaxies. In \cite{sf13,Matos:1998vk} it was found that the SFDM model explains the rotation curves in galaxies, while in \citep{sf40,sf3} it was shown that the model explains the cosmological observations and contains a natural cut-off of the matter power spectrum that allows adjusting the number of satellite galaxies around the host. In \cite{Alcubierre:2001ea} numerical simulations for the collapse of the SFDM were performed and  in \cite{Bernal:2003lwr} it was shown that the density profile of the SFDM is cored. In \cite{Matos:2011pd,Robles:2012kt,Hernandez-Marquez:2026lkw} SFDM with multi-states, finite temperature and symmetry breaking were studied. All of these results were recently corroborated several times with numerical simulations by various groups \citep{sc4,sc10,sc11,sc12,sc13,sc14}.
Almost all DM models that exist so far predict that the distribution of satellite galaxies around large galaxies should be homogeneous \cite{Lipnicky:2016keh} (see \cite{Foot:2016wvj, Berezhiani:2017tth} for some alternatives), with the exception of SFDM, which naturally predicts the correct trajectories of the satellite galaxies \cite{satellite}. This has been one of the great predictions of the SFDM model so far.
SFDM satisfies the Schr\"odinger-Poisson equations, where systems can be in the ground state, or excited states, or both, like atoms. The first non-spherically symmetric excited state is morphologically similar to the 2p state of the Hydrogen atom, it has the shape of the number 8, where spherical harmonic functions with atomic number $l\not=0$ lost their spherically symmetric form. In the case of SFDM, the 8-shaped halo density hijacks the trajectories of the satellite galaxies to align with the polar orbits, breaking the isotropic structure of the satellites around the galaxies, see figure \ref{fig:SchrPois_3d}. Therefore it is capable of explaining the anisotropic distribution of satellite galaxies around their host galaxies, observed in the Milky Way, Andromeda, and Cen A \cite{Bernal:2024wif}. Another relevant feature arises if the scalar field is charged under a local $U(1)$ symmetry, the scalar field is minimally coupled to a gauge field and can act as a source for electromagnetic fields. Charged SFDM configurations have therefore been proposed as a possible mechanism for generating the large-scale magnetic fields observed in galaxies \cite{Hernandez:2018cir,Hernandez-Marquez:2026lkw}. These magnetic fields with strengths of order of $\mu G$ and coherence lengths of kilo-parsecs scales, remain an important astrophysical problem whose origin is not yet fully understood. The possibility that they may arise from a charged SFDM halos provides a natural connection between the internal structure of DM halos and observable electromagnetic phenomena.
This is possible if the excited states of the scalar field are taken into account \citep{Hernandez-Marquez:2026lkw}. 

%
%

{There are several ways to justify the existence of SFDM, for example, in \cite{Hui:2016ltb} the authors propose that SFDM is derived from superstring theory. Here we will adopt the simplest hypothesis that we add the SFDM Lagrangian to the Standard Model (SM), where the scalar field $\Phi$ is in a thermal bath at temperature $T$ endowed with an SFDM potential $V=-m^2\Phi\Phi^*+\lambda/2(\Phi\Phi^*)^2+\lambda/4\Phi\Phi^*T^2+\pi^2/90 T^4$ does not interact with any other SM component. At the beginning of the universe, the SFDM was in thermal equilibrium with the other components of the SM, but decoupled from the SM very early in the universe's history. Like the rest of the components, SFDM cools down in temperature and makes a phase transition at $T_c=2m/\sqrt{\lambda}$, very early in the universe history due to its very tiny self interaction $\lambda$, in order to fulfill the big nucleosynthesis constrains (see \cite{Li:2013nal}). As in the standard model of cosmology, quantum fluctuations going classical collapse the SFDM, which now forms Bose-Einstein condensates to form halos of galaxies, where most of the SFDM particles go to the ground state. Of course, as the collapses occur, the temperature of the halo increases, causing some of the SFDM particles to go into excited states. In \cite{Guzman:2019gqc} it was shown that if most of the particles remain in the ground state, but with a part of the particles in at least one excited state, the system stabilizes and remains as long as the age of the universe, provided that the mass of the SFDM particle is small enough. If not, the system becomes unstable and disappears.  
}

In this work, we use precisely this form of the excited states of the galaxy halo to explain a new observation that has so far been unexplained, the Fermi Bubbles (FBs) \cite{Su:2010qj}. {The Fermi Bubbles are two approximately symmetric, lobe-shaped gamma-ray-emitting structures extending above and below the Galactic disk of the Milky Way, roughly along its rotation axis, to distances of about 9 kpc. Despite their discovery in 2010, their origin and formation history remain unclear. Leading proposed scenarios include outflows driven by past AGN activity and supernova feedback \cite{galaxies6010029}.} 

Here we investigate a different possibility: that the Fermi Bubbles are associated with an excited SFDM configuration and investigate whether it can provide the energy required to account for the observed Fermi Bubbles. This scenario establishes a possible connection between the quantum structure of the DM halo, galactic magnetic fields, and the high-energy emission observed in the FBs region.\\ 
\textbf{The model:}
We consider an epoch after spontaneous symmetry breaking, when the fluctuation $\delta\Phi$ that forms the galaxy has already collapsed. At this stage, the Lagrangian is given by
\begin{eqnarray}\label{eq:Lagrangian}
\mathcal{L}&=&-(\nabla_\mu\Phi+iqB_\mu\Phi)(\nabla^\mu\Phi^*-iqB^\mu\Phi^*)-m_{\Phi}^2\Phi\Phi^*\nonumber\\
&-&\frac{\lambda}{2}(\Phi\Phi^*)^2-\frac{1}{4}B_{\mu\nu}B^{\mu\nu}
\end{eqnarray}
where $B_\mu$ is the  gauge field
of the SFDM, with fundamental charge $q$ and Faraday tensor $B_{\mu\nu}=B_{\mu;\nu}-B_{\nu;\mu}$, $m_{\Phi}:=m^2-(\lambda/2)T^2$ is the effective mass of the scalar field.
The scalar field $\Phi=\Phi_0(\eta)+\delta\Phi(x^i)$  is the dark matter of the universe and has the dominant part of the gravitational field. The background scalar field $\Phi_0$ fulfills the Klein-Gordon equation and it was recently shown that the perturbation $\delta\Phi$ gives rise to the formation of an halo whose density profile is given by $\rho\sim|\delta\Phi\delta\Phi^*|$ and whose final collapse is a halo in the form of
an atom, with a first excited state in the form of two lobes along the north-south direction \cite{Bernal:2024wif}. The SFDM density profile looks like the one shown on the left side of figure \ref{fig:SchrPois_3d}. This reminds us very well the shape of the FBs in the Milky Way.\\
Furthermore, in \cite{Hernandez-Marquez:2026lkw} it was found that if the gauge field $B_{\mu}$ is treated as a perturbation and is the Standard Model photon and if $q\sim 10^{-48}e$, then the multiscalar dark matter can generate electric and galactic magnetic fields with intensities like the observed $\mu G$. These galactic magnetic fields are given in terms of spherical Bessel functions, the galactic magnetic field generated by this charged SFDM for the Milky Way is shown in the upper right panel of Figure \ref{fig:SchrPois_3d}. This galactic magnetic field was obtained using the values for the
free parameters of the model reported in \cite{Hernandez:2018cir} and its
strength decreases as the
radius increases.\\
Then, if we model the Fermi Bubbles as two spheres with radius of $7$ Kpc centered at $(0,0,7)$ Kpc and $(0,0,-7)$ Kpc with respect to the galactic plane of the Milky Way, the total electromagnetic energy contained within these regions is given by: 
\begin{equation}\label{eq:Maxwell}
E_{EM}=\iiint_{V_1 \cup V_2}\dfrac{B^2+E^2}{8\pi}\,dV,
\end{equation}
where $V_1$ and $V_2$ are the respectively volume of the FBs, enclosed within these to spheres. Using the galactic electromagnetic fields predicted by \cite{Hernandez-Marquez:2026lkw},
 then we obtain $E_B=2.6193\times10^{66}$ eV and $E_E=7.4311\times10^{57}$ eV at $z=0$, $E_B=4.4359\times10^{71}$ eV and $E_E=1.3983\times 10^{64}$ eV at $z=10$ and $E_B=2.6193\times10^{81}$ eV and $E_E=7.4313\times 10^{75}$ eV at $z=1100$. Therefore, the energy of the Fermi Bubbles is between $E_{FB}=6.2415\times10^{66}$ eV to $E_{FB}=6.2415\times 10^{68}$ eV. These results indicate that, because the electromagnetic energy density decreases as the Universe expands, a substantially larger electromagnetic energy reservoir was available at earlier cosmic times. {Consequently, our model provides sufficient electromagnetic energy to power several non-thermical processes, including:  Fermi I acceleration, Fermi II acceleration and the local electric fields generated in the current sheets because of magnetic reconnection through the conservation electromagnetic energy density:
\begin{equation}\label{eq:Maxwell}
\dfrac{\partial}{\partial t}(\dfrac{B^2+E^2}{8\pi})+\dfrac{c}{4\pi}\nabla \cdot (\vec E \times \vec B)=-\dfrac{1}{c}\vec V \cdot(\vec J\times \vec B)-\dfrac{J^2}{\sigma}
\end{equation}
where $\vec V$ is the fluid velocity vector field and $\sigma$ is the electrical conductivity. These mechanisms can accelerate electrons while simultaneously driving magnetohydrodynamic (MHD) instabilities that generate turbulence. Through the MHD cascade, energy is progressively transferred to smaller scales, where it can interact with charged particles and accelerate a small fraction of them to relativistic energies \cite{Shukurov2021}. Over time, a positive feedback mechanism is established between turbulence and the magnetic fields, allowing these non-thermal processes to be sustained even after the electromagnetic energy directly supplied by the electromagnetic fields generated by the SFDM becomes insufficient. Consequently, the free electrons in the Milky Way can be continuously accelerated to relativistic energies through the energy transferred from the electromagnetic field via the aforementioned mechanisms. 

%
%

\begin{figure*}[t]
\centering
\includegraphics[width=8cm]{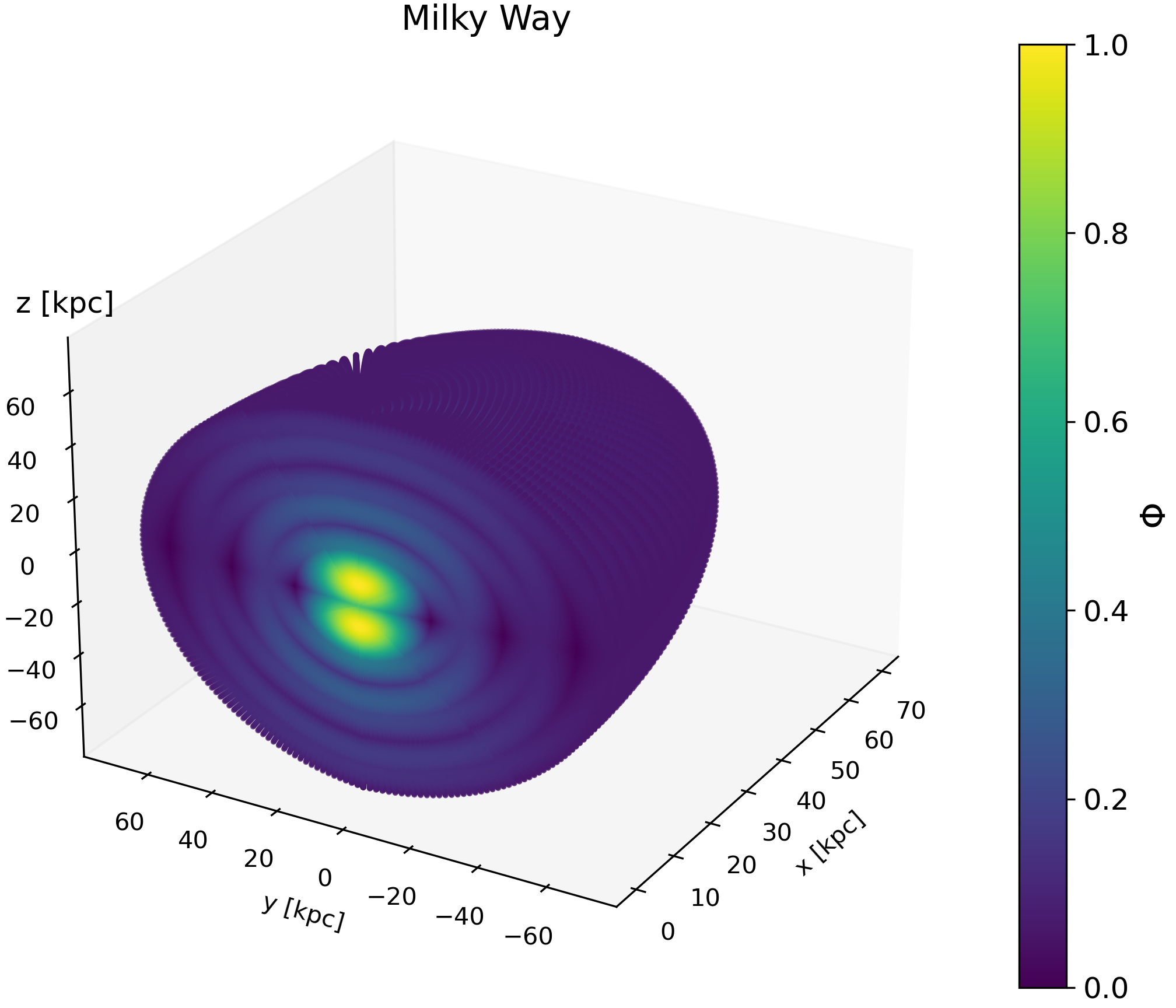}\hspace{0.5cm}\hspace{0.5cm}
\includegraphics[width=8cm]{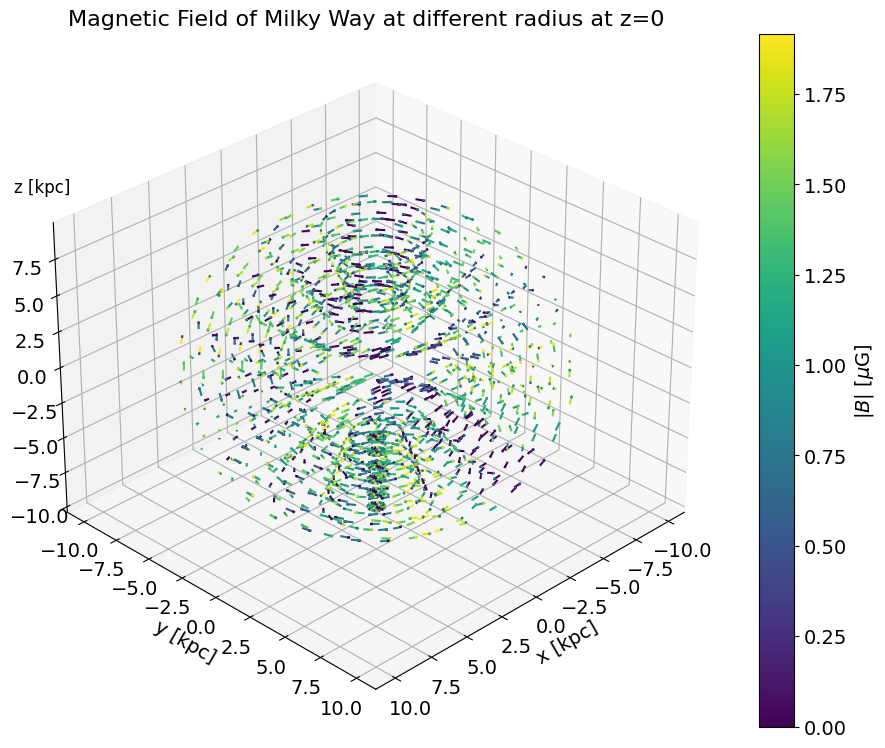}\\
\includegraphics[width=8cm]{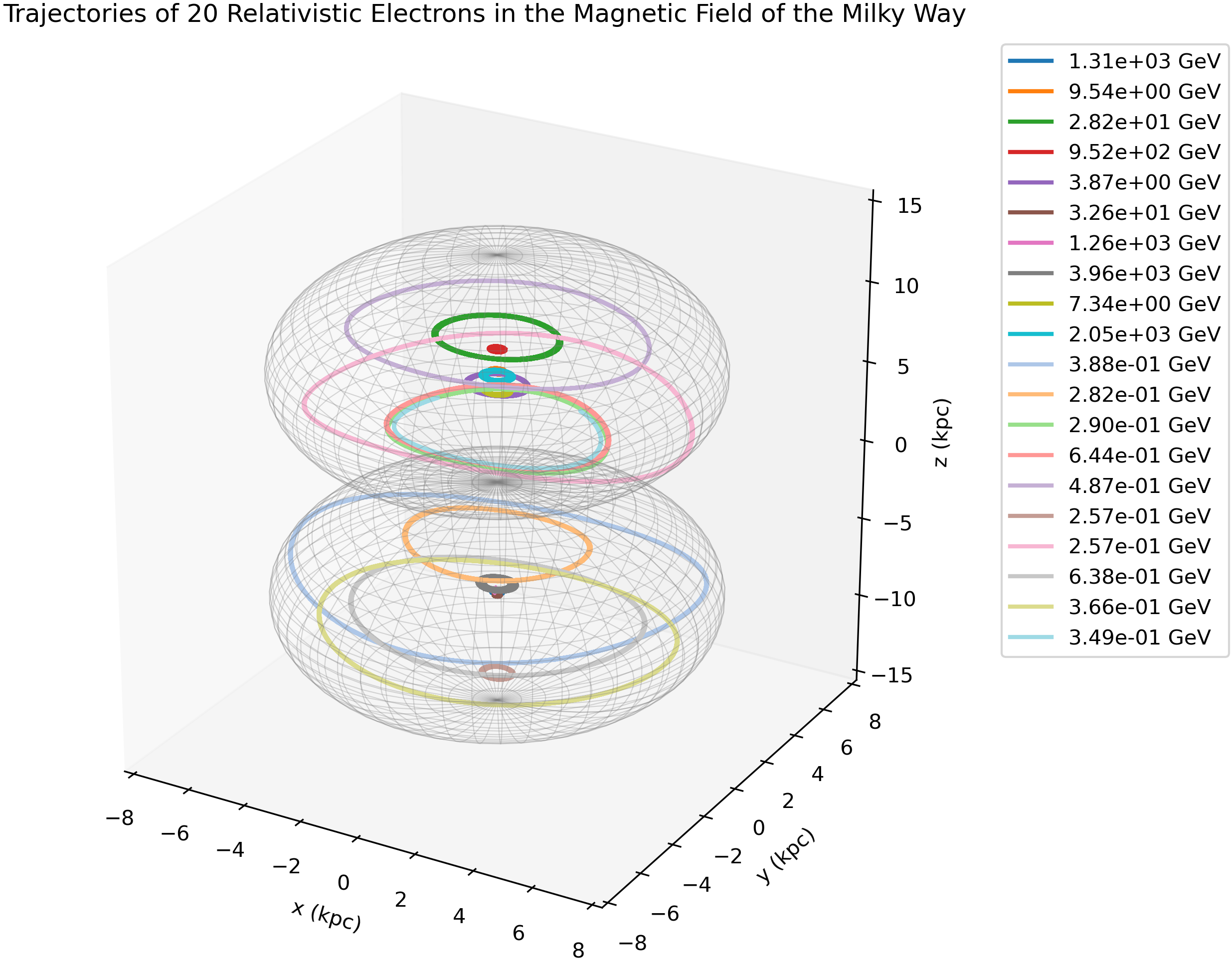} \hspace{0.5cm}\includegraphics[width=8cm]{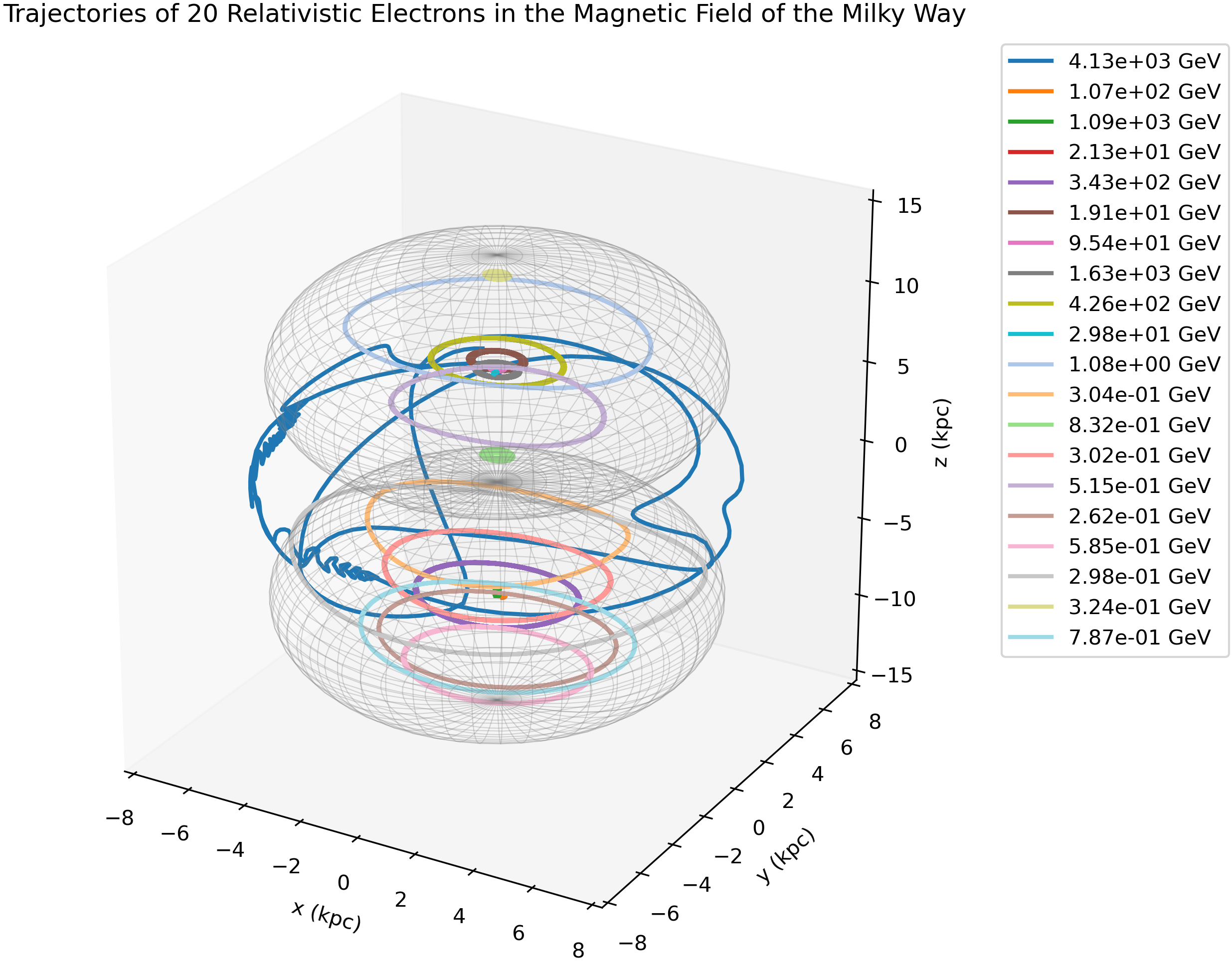}
\caption{\textbf{Upper right panel:} Two combined states of the scalar field: the ground state and the first excited state of the Milky Way, which is similar for other galaxies. \textbf{Upper left panel:} Milky Way's magnetic field obtained from \cite{Hernandez-Marquez:2026lkw} using rotation curves to fix the model's free parameters and the magnetic field constraints. \textbf{Lower left panel:} The trajectories of twenty relativistic electrons following inverse Compton scattering with CMB photons. The electron energies, ranged from 255 MeV to 8 TeV after the collision, and positions within the Fermi Bubbles region were randomly assigned. The legend indicates the electron energies, and the simulation spans approximately two million years. \textbf{Lower right panel:} Trajectories of another set of twenty relativistic electrons, shown to better visualize particle confinement within the Fermi Bubbles region.}
\label{fig:SchrPois_3d}
\end{figure*}
\begin{figure*}[t]
\centering
\begin{minipage}[t]{0.48\textwidth}
    \centering
    \includegraphics[height=7cm]{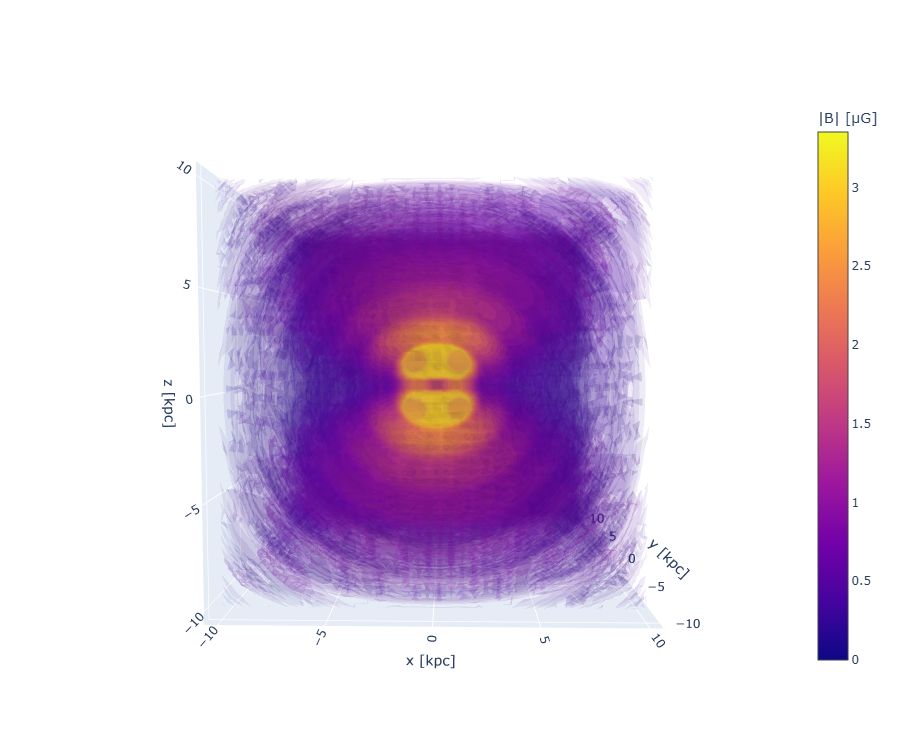}
\end{minipage}
\hfill
\begin{minipage}[t]{0.48\textwidth}
    \centering
    \includegraphics[height=7cm]{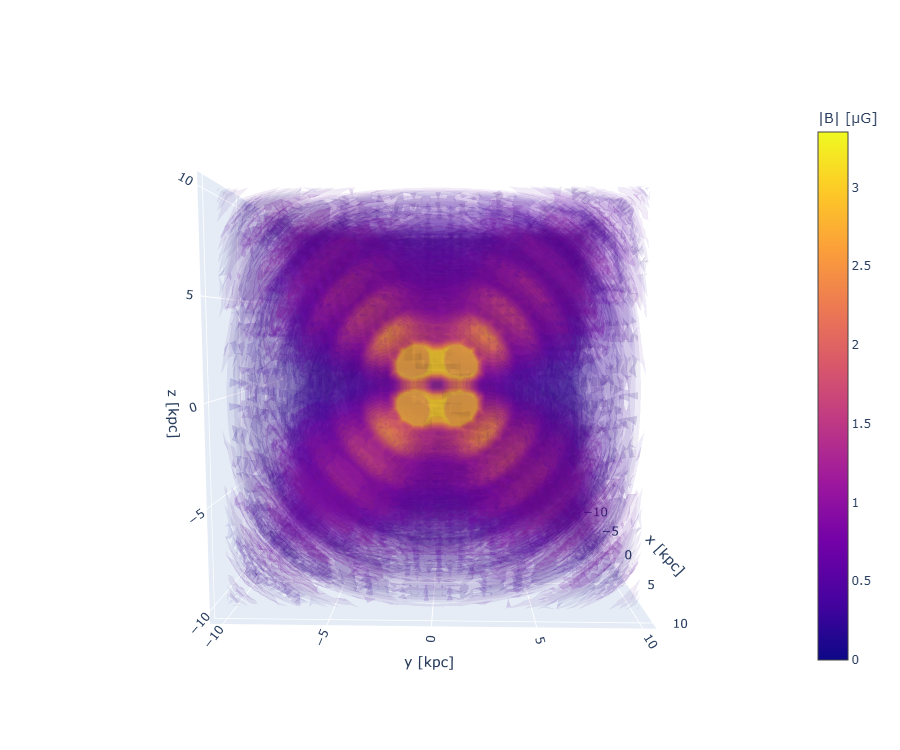}
\end{minipage}
\caption{Left panel: $3D$ map of the magnetic field strength viewed from the x-z plane.
Right panel: $3D$ map of the magnetic field strength viewed from the y-z plane. This figure resembles the X-shaped structure observed in the magnetic fields of galaxies; see for example \cite{Soida2011,Heesen:2009sg, Haverkorn2011,Ferriere2014}}.
\label{fig:strength}
\end{figure*}
\begin{figure*}[t]
\centering
\begin{minipage}[t]{0.48\textwidth}
    \centering
    \includegraphics[height=6cm]{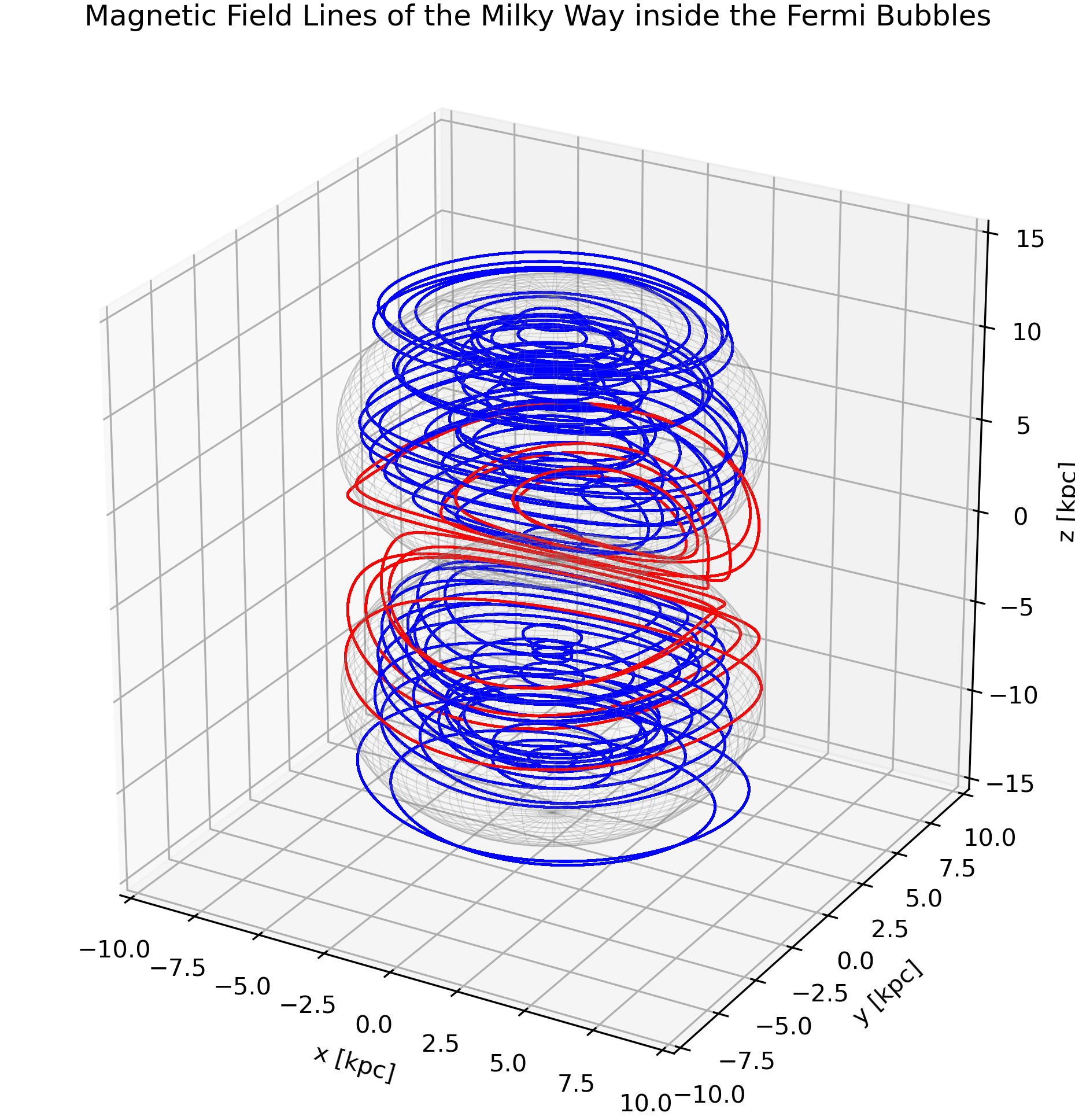}
\end{minipage}
\hfill
\begin{minipage}[t]{0.48\textwidth}
    \centering
    \includegraphics[height=6cm, width=6cm]{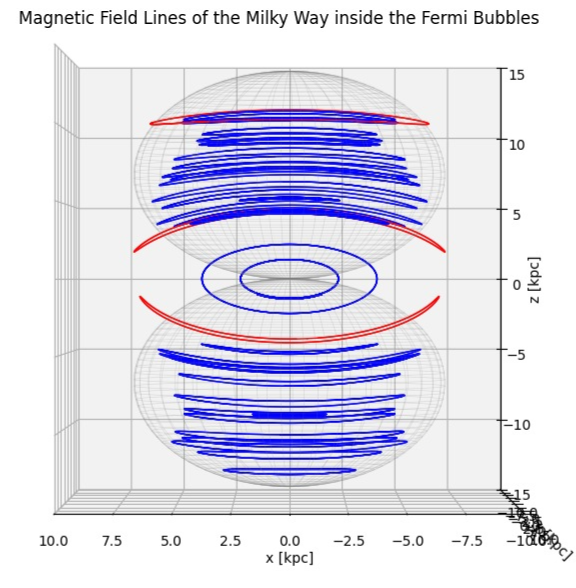}
\end{minipage}
\caption{Left panel: Magnetic field lines of the Milky Way within the Fermi Bubbles. The field lines are traced over around two millions years. A total of 160 field lines are plotted by solving $\dfrac{d\vec r}{ds}=\pm \dfrac{\vec B}{|B|}$, where $s$ denotes the path length along the magnetic field. The blue lines represent magnetic field lines that originate within the Fermi Bubbles and remain confined inside this region after propagating over the full two million years. In contrast, the red lines correspond to magnetic field lines that also originate within the Fermi Bubbles but escape from them at some point along their trajectories. Among the 160 traced field lines, $13.75\%$ escape from the Fermi Bubbles, while the remaining 86.25\% remain confined within the structure. This high degree of confinement indicates that the transport of relativistic particles is strongly anisotropic, occurring predominantly along the direction of the Galactic magnetic field lines. Right panel: Magnetic field lines viewed from the x-z plane, showing loops around the y-axis. These field lines can therefore guide charged particles toward the Fermi Bubbles region}
\label{fig:combined}
\end{figure*}
 Now, we consider that the high-energy photons observed in the
Fermi Bubbles are produced through inverse Compton
scattering between these relativistic electrons and CMB photons, which typically have energies of
approximately $10^{-3}$ eV. Since the photon energy is much smaller than the electron rest energy, the interaction occurs in the Thomson regime. Therefore, the energy gained
by the photons depends strongly on the velocity of the electrons through the factor $\gamma^2$: 
\begin{equation}
\label{eq:Maxwell}
E_f\approx4\gamma^2E_i,
\end{equation}
where $E_f$ is the acquired energy of the photon and $E_i$ is its initial value, $\gamma$ is the Lorentz factor of the electron. Then, in order to obtain photons with energies of MeV and TeV, we consider ultra-relativistic electrons with $\gamma\sim 10^{2}-10^{7}$. From this, we find that the electrons responsible for inverse Compton scattering with CMB photons have energies ranging from 255 MeV to 8 TeV.\\
We then calculate the trajectories of electrons with energies within this range as they propagate through the Fermi Bubbles by solving the Lorentz force equation for charged particles moving in a magnetic field
\begin{equation}\label{eq:Maxwell}
\gamma m_e\dfrac{d\vec v}{dt}=-e\vec v\times \vec B,
\end{equation}
where $\vec{v}$ is the electron's velocity, $e$ is the electron charge and $\vec{B}$ is the Milky Way (MW) magnetic field generated by the gauge field $B_{\mu}$, obtained in \cite{Hernandez-Marquez:2026lkw}. 
To integrate the trajectories of the electrons the relativistic Boris pusher was employed since it is one of the most widely used, thoroughly validated, and reliable numerical algorithms for solving the relativistic Lorentz equation of motion. Originally introduced by J. P. Boris in 1970  for Particle-in-Cell (PIC) plasma simulations, it has since become the standard method for advancing charged particles in electromagnetic fields \cite{Boris1970}.

\textbf{Results}:
In the lower-right and lower-left panel of Figure \ref{fig:SchrPois_3d}, we show the trajectories of 20 ultra-relativistic electrons with energies  in the ranges ($2.57\times 10^{-1}-3.96\times 10^3$ GeV) and $2.62\times 10^{-1}-4.13\times 10^{3}$ GeV, respectively,as they propagate through the Fermi Bubbles under the influence of the galactic magnetic field generated by the charged SFDM. The initial positions of the electrons were randomly assigned. Due to the presence of the galactic magnetic field, the electrons remain confined within the Fermi Bubbles for approximately two million years. During this time, the electrons undergo inverse Compton scattering with CMB photons, transferring energy to them and producing high-energy photons. The resulting photon energies span the range $E_f \sim 1$ keV to 1 TeV. In the left panel of Figure \ref{fig:strength} it is shown a three-dimensional map of the Milky Way magnetic field strength viewed in the ($x-z$) plane while the right panel shows the corresponding field strength viewed from the ($y-z$) plane. This figure resembles the $X$-shaped magnetic field configuration observed in galaxies \cite{Soida2011,Heesen:2009sg,Haverkorn2011,Ferriere2014}. The left panel of Figure \ref{fig:strength} shows $160$ magnetic field lines that originate within the Fermi Bubbles. The field lines shown in blue remain confined within the Fermi Bubbles, whereas those shown in red escape from the Bubbles at some point. Approximately, $86.25\%$ of the field lines remain confined within the structure of the Fermi Bubbles. Therefore, this suggests that the transport of relativistic particles is strongly anisotropic, occurring predominantly
along the direction of the Galactic magnetic field-lines. In the right panel we show the magnetic field lines viewed from the $x-z$ plane, showing loops around the $y-$ axis, indicating that the magnetic field can guide charged particles toward the region of the Fermi Bubbles.\\
Finally, if charged SFDM constitutes the dark matter of the Universe and is responsible for generating galactic magnetic fields, we would expect to observe Fermi Bubbles in other galaxies, particularly in large galaxies. The number of observable events depends on the geometry of the inverse Compton scattering process. Thus, only a fraction of the resulting photons may fall within the energy range accessible to current detectors, while the remaining photons may either have energies below the detection threshold or extend beyond the energy range of current instruments.\\
\textbf{Conclusions}: In conclusion, because of the quantum character of SFDM, this is the only DM model that allows predicting a DM density profile with bubbles, as in atoms. With this axially symmetric shape of the profile, the SFDM model is able to explain not only the rotation curves in the galaxies but also the observed anomalous behavior of the satellite galaxies, the VPOS, in a very natural way \cite{satellite}. Now, in this work, we show that this DM profile can explain the FB observed in galaxies, provided that the SFDM has an extremely small charge $q\sim 10^{-48}e$, consistent with currents observational bounds. It is extremely interesting that the SFDM model can explain in such a natural way so many observable features of the galaxies and the cosmos, without further physical complications, even such unexpected observations as the VPOS or the FB. If this hypothesis is correct, we should see FBs in more large galaxies and gamma rays at higher energies in the near future.

\textbf{Acknowledgments.}
This work was partially supported by SECIHTI Mexico under grants CBF-2025-G-1720 and CBF-2025-G-176. MH-M acknowledges financial support from SECIHTI postdoctoral fellowship.

\bibliographystyle{unsrt}
\bibliography{references} 

\end{document}